\documentclass[a4paper,12pt]{article}
\usepackage[utf8]{inputenc}
\usepackage{color,soul}
\usepackage{authblk}
\usepackage{float}
\usepackage{graphicx}
\graphicspath{ {./main figures/} }

\usepackage{amsmath,amssymb}

\usepackage{setspace}
\usepackage{float}
\floatstyle{plaintop}
\restylefloat{table}

\usepackage[square,numbers,sort&compress]{natbib}
\usepackage{hyperref}
\usepackage[font=footnotesize,figurename=Fig.]{caption}
\usepackage[labelfont=bf]{caption}

\usepackage[margin=2.0cm]{geometry}

\title{\vspace{-2cm}\LARGE 
Structural Diversity in Electrohydrodynamically Driven Active and Organized Liquid States}
\author[1]{\vspace{+0.0cm} \normalsize Geet Raju}
\author[1]{\normalsize Nikos Kyriakopoulos} 
\author[1]{\normalsize Jaakko V. I. Timonen*\vspace{+0.0cm}}
\affil[1]{\small Department of Applied Physics, Aalto University School of Science, P.O. Box 15100, Espoo FI-02150, Finland\vspace{-1cm}}
\date{}

\begin{document}

\renewcommand{\figurename}{\textbf{Figure}}

\maketitle

\noindent\textbf{Spontaneous emergence of organized states in materials driven by non-equilibrium conditions is of significant fundamental and technological interest.\cite{Cross2009} In many cases, the organized states are complex, hence, with some well-studied exceptions, their emergence is challenging to predict.\cite{Cross2009,Marchetti2013,Fialkowski2006,Fodor2016} In this article, we show that an unexpectedly diverse collection of dissipative organized states can emerge in a simple biphasic system consisting of two liquids under planar confinement. We drive the liquid-liquid interface, which is held together by capillary forces,\cite{deGennes2004} out of thermodynamic equilibrium using DC electrohydrodynamic shearing.\cite{Vlahovska2019} As a result, the interface goes through multiple spontaneous symmetry breakings, leading to various organized non-equilibrium states. First, at low shearing, the shear-deformed interface becomes unstable and a 1D quasi-static corrugation pattern emerges. At slightly higher shearing, we observe topological changes that lead to emergence of active self-propulsive fluidic filaments and filament networks, as well as ordered bicontinuous fluidic lattices. Finally, the system transitions into active self-propulsive droplets, quasi-stationary dissipating polygonal and toroidal droplets, and ultimately to a chaotic active emulsion of non-coalescing droplets with complex interactions. Interestingly, this single system captures many features from continuum non-equilibrium pattern formation\cite{Cross2009} and discrete active particles,\cite{Marchetti2013} which are often considered separate fields of study. The diversity of observed dissipative organized states is exceptional and points towards many new avenues in the study of electrohydrodynamics,\cite{Vlahovska2019} capillary phenomena,\cite{deGennes2004} non-equilibrium pattern formation,\cite{Cross2009} and active materials.\cite{Marchetti2013,Hagan2016}} 

\newpage
\subsection*{Introduction}
Spontaneous organization of matter into complex structures under non-equilibrium conditions is of both major fundamental and technological interest across disciplines, from physics to biology.\cite{Cross2009,Whitesides2002,Marchetti2013} In the context of physics, it is often manifested by symmetry breaking and emergence of unexpected patterns in continuum systems driven by external fields,\cite{Cross2009} and as structural organization of externally driven\cite{Grzybowski2000} and self-propulsive (active) particles.\cite{Palacci2013} In (bio)chemical systems, it is often seen as active gelation\cite{Prost2015} of both biological\cite{Nedelec1997} and synthetic molecules.\cite{Boekhoven2015,vanRossum2017,De2018} In biological systems, it is manifested, for example, by collective motion of microscopic\cite{Marchetti2013} and macroscopic\cite{Jhawar2020,Couzin2003} organisms. Although progress has been made towards theoretical descriptions,\cite{Cross2009,Marchetti2013,Fialkowski2006,Fodor2016} predicting the systems and required non-equilibrium conditions that lead to the emergence of complex structures remains still a challenge.

There is a rich literature on using electrohydrodynamics (EHD)\cite{Vlahovska2019} to drive agent motion, be it solid particle rotation and translation, or 3-dimensional liquid droplet rotation and deformation. Significant advances has been made towards understanding the properties of such systems regarding their interactions and collective behavior, in the case of solid spheres,\cite{Bricard2013,Bricard2015,Morin2017, Rozynek2014} and their deformation and splitting, in the case of liquid droplets.\cite{Dommersnes2013,Ouriemi2014,Ouriemi2015, Salipante2010,Vlahovska2016,Brosseau2017,Varshney2012,Tadavani2016,Timonen2013} Even so, there are still many areas left unexplored: the shapes of the driven agents have always been symmetric (e.g. spheres); shapes of lower symmetries and the changes they induce to the local and global EHD flows that govern the behavior of the system have not been explored in depth. In addition, while some examples of confinement have been explored in the case of solid particles, the geometry has never been confining in the case of liquid droplets, thus the droplets have been allowed to move and expand freely in all three dimensions. In this article, we confine a simple biphasic liquid system between two non-wetting surfaces, thus forcing a quasi-2-dimensional behavior. The confinement, along with the deformability of the liquid drops and their material properties, lead to strong deformation of the interface when an external electric field is introduced. The deformation is driven by nontrivial EHD flows and shearing of the interface. As a result, an unexpected diversity of complex, non-equilibrium fluidic patterns and fluidic objects of various shapes emerge (Fig. 1, Movie S1). Our system relies on a novel combination of two immiscible oils: fluorinated perfluoropolyether (PFPE) and a dodecane (DD) solution containing 75 or 150 mM bis(2-ethylhexyl) sulfosuccinate sodium salt (Aerosol OT, AOT), confined between two parallel glass slides coated with indium tin oxide (ITO) and separated by a distance $h$ (Fig. 1A, Supplementary Fig. S1), between 36 and 68 $\mu$m (Supplementary Tables S1-2).

The combination of oils used has three critical properties. Firstly, the oils have a significant conductivity contrast ($\sigma_{\mbox{\scriptsize{PFPE}}}\approx 10^{-17}$ S/m $\ll$ $\sigma_{\mbox{\scriptsize{150 mM AOT/DD}}} \approx 10^{-8}$ S/m, Supplementary Table S3) that is required for driving strong electrohydrodynamic flows near the oil-oil interface.\cite{Vlahovska2019} Secondly, the contact angle $\theta$ of the essentially insulating PFPE on the ITO surfaces (used as electrodes) surrounded by the AOT/DD phase is close to ideal nonwetting, i.e. $\theta \approx 180^\circ$. Thirdly, the adhesion of the PFPE phase on the ITO is negligible. The latter two properties allow the PFPE phase to form highly mobile slab-shaped droplets with semicircular oil-oil interface when confined in a planar cell between two parallel ITO-coated electrodes in thermodynamic equilibrium (Fig. 1A)\cite{deGennes2004}. We drive this interface out of the equilibrium by applying a voltage $U$ between the ITO-coated electrodes, resulting in electric field $E_0=U/h$. This leads to a dynamic charge density at the oil-oil interface that experiences a Coulombic destabilizing (shearing) force (Fig. 1A),\cite{Vlahovska2019} that is counteracted by the interfacial tension $\gamma\approx8$ mN/m (see Materials and Methods in Supplementary Materials).\cite{deGennes2004} At suitable balance between these forces, the interface (Fig. 1B) goes through numerous unexpected symmetry breakings and changes in topology that lead to a striking variety of non-equilibrium patterns and active objects (Fig. 1C): dissipative quasi-1D patterns at the oil-oil interface, self-propulsive (active) liquid filaments and dissipative filament networks and lattices, and self-propulsive and electrohydrodynamically sculpted droplets.

\begin{figure}[t]
\includegraphics[width=\linewidth]{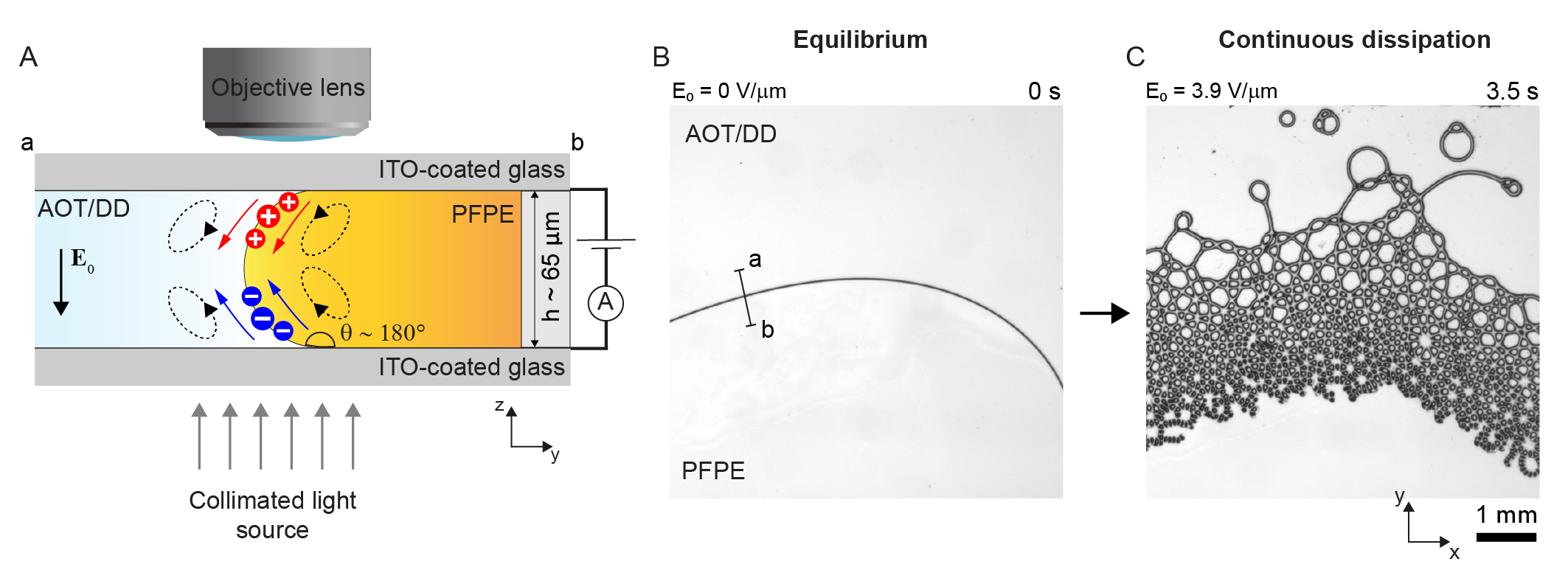}
\caption{
\textbf{Emergence of diverse non-equilibrium fluidic structures in a simple electrohydrodynamically driven planar system.} 
\textbf{A,} Scheme of the cross-section of a perfluoropolyether (PFPE) drop (orange) surrounded by a mixture of dodecane and AOT (AOT/DD) (light blue) between two ITO-coated glass slides. Red and blue arrows indicate Coulombic force and flow direction at the oil-oil interface for positive (red) and negative (blue) interfacial charges. Dashed black loops and arrows indicate resulting flow fields outside the interface. The same color scheme and notation is used throughout the article.
\textbf{B,} An image of the oil-oil interface in equilibrium state observed in the experiments.
\textbf{C,} An image of a typical non-equilibrium state comprising active filaments, random filament networks and ordered 2D fluidic lattices, and active and self-propulsive electrohydrodynamically sculpted droplets and active emulsions.}
\label{fig:overview}
\end{figure}

\subsection*{Interfacial instabilities and patterns}
When the electrohydrodynamic driving is weak (small electric field magnitudes), the interfacial tension largely keeps the oil-oil interface close to the equilibrium shape. The only small indication of the electrohydrodynamic shearing comes from the minute reshaping of the interface upon increasing the electric field. This is observed in our microscopy experiments (see Materials and Methods) as changes in the dark band corresponding to the oil-oil interface created by the refraction of light from the curved interface (Fig. 2A, Supplementary Fig. S2A). When the voltage is increased, the width of the dark band gradually increases from the equilibrium value $w_0\approx h/2$ (as expected for semicircular interface) as $w = w_0 \left( 1+ A E_0^\alpha \right)$, where $\alpha \approx 2.6$ (Fig. 2B, Supplementary Fig. S2B-C). The widening is driven by the anti-symmetric charging of the oil-oil interface with respect to the mid-plane of the cell and the resulting symmetric electrohydrodynamic shearing of the interface towards the mid-plane and the apex of the interface (Fig. 2A). This is analogous to the Taylor flow pattern observed in freely suspended, non-confined spherical droplets, which leads to axisymmetric droplet flattening, known since the 1960s.\cite{Taylor1966} The widening of the interface is reversible, and the oil-oil interface returns to the equilibrium state when the electric field is turned off (Supplementary Fig. S2D).

At slightly larger electric fields above a threshold value $E_c \approx$ 4 V/$\mu$m, the shear-widened interface (Fig. 2A) is no longer stable, and thus undergoes a spontaneous symmetry breaking. This leads to the population of the oil-oil interface with periodically alternating peaks and valleys (Fig. 2D, Movie S2) in a pattern reminiscent of Faraday waves\cite{Westra2003} and the Rosensweig pattern,\cite{Rosensweig2014} but now in one dimension. If the applied field is only slightly above $E_c$ (i.e. close to the threshold), the instability and pattern formation takes place in a few tens of milliseconds in two distinct steps (Fig. 2C): first the interface elongates (taking ca. 1-2 milliseconds), followed by a short quiescent time and the spontaneous symmetry breaking (taking ca. 5 milliseconds). If the applied field is significantly above threshold $E_c$, the two processes are not well separated in time. The steady-state spacing between the peaks, $\lambda_{pp}\approx$ 65 $\mu$m, is close to the cell height $h=68 \pm 0.6$ $\mu$m (Fig. 2D). However, careful microscopic observation shows that the peaks appear in pairs with alternating light and dark valleys between them (Fig. 2D), indicating that the true periodicity is $\lambda \approx 2 h$. This observation is reinforced by the intensity profile analysis along lines normal to the interface at neighboring peak and valley positions (Fig. 2E), showing different profiles for the valleys. When the imaging system is focused at the mid-plane of the cell, the peaks appear to be in focus, while the valleys are out of focus. This suggests that the edges of the valleys are alternating above and below the mid-plane of the cell (Fig. 2D). In addition, experiments using high-speed microscopic imaging show that tracer particles (microscopic PFPE droplets) circulate near every second valley in a plane perpendicular to the interface, completing a full loop in $\tau\approx$ 8 ms (Movie S2). These observations point towards a mechanistic picture wherein the symmetric Taylor-like electrohydrodynamic flow is spontaneously broken and the system adopts a state similar to the rotational Quincke flow,\cite{Quincke1896} only with a periodically varying direction of rotation along the oil-oil interface (Fig. 2E-F).

\begin{figure}[t]
 \centering
\includegraphics[width=1.00\columnwidth]{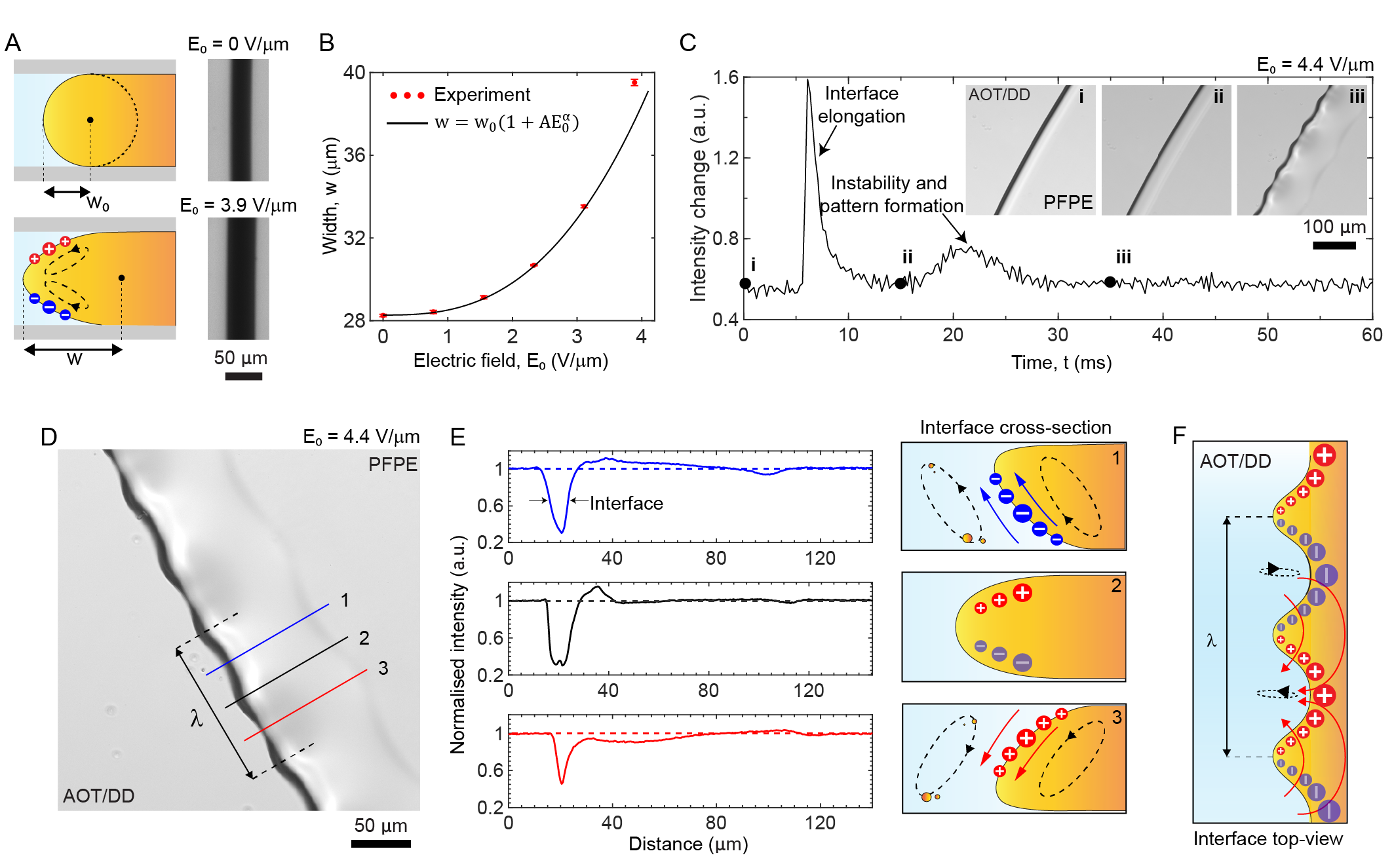}
\caption{
\textbf{Electrohydrodynamically driven reshaping of the oil-oil interface and emergence of dissipative interfacial patterns.} 
\textbf{A,} Schemes and images of the oil-oil interface, with and without an applied field.
\textbf{B,} Width (FWHM) of the oil-oil interface as a function of electric field magnitude strength. The solid line denotes the fit $w = w_0 \left( 1+ A E_0^{\alpha} \right)$, with $w_0=28.25$, $A=0.0097$ and $\alpha = 2.6$. The rightmost experimental point was ignored in the fit, as the corrugation pattern starts appearing near that field magnitude.
\textbf{C,} Change in image intensity between consecutive frames as a function of time ($\Delta I(t) = \sum\sum |I_{ij}(t+\Delta t)-I_{ij}(t)|$, where $I_{ij}(t)$ is the matrix of pixel intensities in the frame recorded at time $t$) when an electric field above $E_c$ is applied (Movie S2). Insets show images before voltage is applied, after the interface has elongated, and after the elongated interface has developed the pattern.
\textbf{D,} An image showing the steady-state dissipative interfacial pattern.
\textbf{E,} Normalised intensity of the interfacial patterns along the lines marked in \textbf{D} and the corresponding cross-sectional schemes at the interface.
\textbf{F,} Top-view scheme of the interfacial pattern.}
\label{fig:interface}
\end{figure}

\subsection*{Active filaments and bicontinuous dissipative fluidic lattices}
Further increasing the strength of electrohydrodynamic driving leads to the emergence of another class of patterns and structures that are no longer spatially restricted to the vicinity of the original oil-oil interface. Rather, the forming patterns and structures emerge from the oil-oil interface (Fig. 3A) and propagate to fill the whole sample cell (Movie S3). The formation of these patterns and structures is enabled by topological changes in the system, wherein the electrohydrodynamic shearing punches holes through the slab-like PFPE drop (Fig. 3B, Movie S4). The holes grow rapidly, and if the expansion of the holes is unrestricted, active and self-propulsive filaments are produced (Fig. 3C-G).

The observed active filaments (Fig. 3C-G) are quasi-self-propulsive objects with a diameter close to the cell height. Tracer particles circulate rapidly around the filaments, typically one rotation in  $\tau\approx$ 1 ms (Movie S5). This suggests that the filaments are in a state of Quincke rotation and that the propulsion originates from slightly asymmetric electrohydrodynamically driven hovering of the filaments between the confining surfaces.\cite{Pradillo2019} Interestingly, the active filaments also form complicated filament networks by looping back onto themselves or coupling with other filaments (Fig. 3F-G, Movie S6). These structures exhibit frustration in the rotational flow direction at the junctions, which leads to the formation of small bumps (Fig. 3E-G) and optical contrast at the locations where the tangential flows vanish. This is consistent with the conclusion regarding the flow fields near the peaks at the oil-oil interface at lower voltages (Fig. 2E-F).

\begin{figure}[h!]
\includegraphics[width=\linewidth]{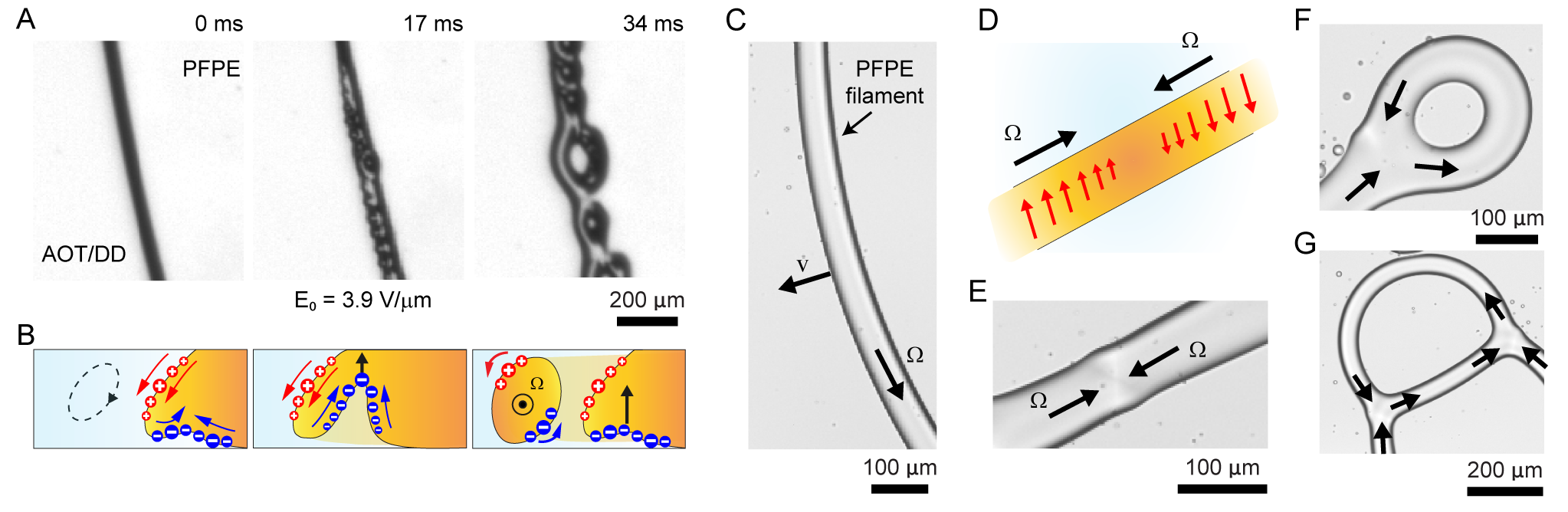}
\caption{
\textbf{Emergence of active filaments and filament networks.}
\textbf{A,} Images of the oil-oil interface at zero field and upon application of voltage $E_0>E_c$ (Movie S3).
\textbf{B,} Scheme illustrating the mechanism for emergence of active filaments. 
\textbf{C} Image of a simple active filament with rotational (Quincke) flow resulting in a translational velocity $v \approx 3.5$ mm/s (Movie S5).
\textbf{D,} Scheme of an active filament illustrating the emergence of frustration-induced bumps.
\textbf{E,} A corresponding image of an active filament with said bump. 
\textbf{F-G,} Images of typical filament junctions with frustration-induced bumps indicated with black arrows (Movie S6).
}  
\label{fig:tiling}
\end{figure}

On the other hand, if the hole growth is limited, dissipative ordered fluidic lattices appear (Fig. 4). The observed dissipative fluidic lattices (Fig. 4A, E) are quasi-static, bicontinuous and periodic fluidic structures with lattice constants on the same order of magnitude as the cell height. Dissipative lattices form via the repeated punching of holes through the PFPE droplet under conditions where the hole growth is highly restricted (Fig. 4A-B). Two lattices were observed to be especially frequent and stable: a trihexagonal lattice (Kagome lattice, Fig. 4E-F, Movie S7) and a double square lattice (Fig. 4A-D, Movie S8). In both lattice types, the neighboring lattice sites (pockets of AOT/DD in the PFPE slab) exhibit liquid flow in opposing directions, either upwards or downwards (Fig. 4B). This is seen directly as circulatory flows between neighboring lattice sites in tracer particle trajectories (Fig. 4C, Movie S8), which again suggests a Quincke-like state of flow\cite{Quincke1896}, but now in the form of a periodic fluidic lattice. It appears that only some lattice symmetries are allowed, due to the requirement of neighboring sites to flow in opposite directions. This is in agreement with both the two observed lattices and in the lack of observation of the typical simple hexagonal tiling.

The fluidic filaments and lattices are reversible: when the voltage is turned off, the structures relax back to simple unstructured droplets (Movie S9) that eventually coalesce back to a single PFPE droplet.


\begin{figure}[h!]
\includegraphics[width=\linewidth]{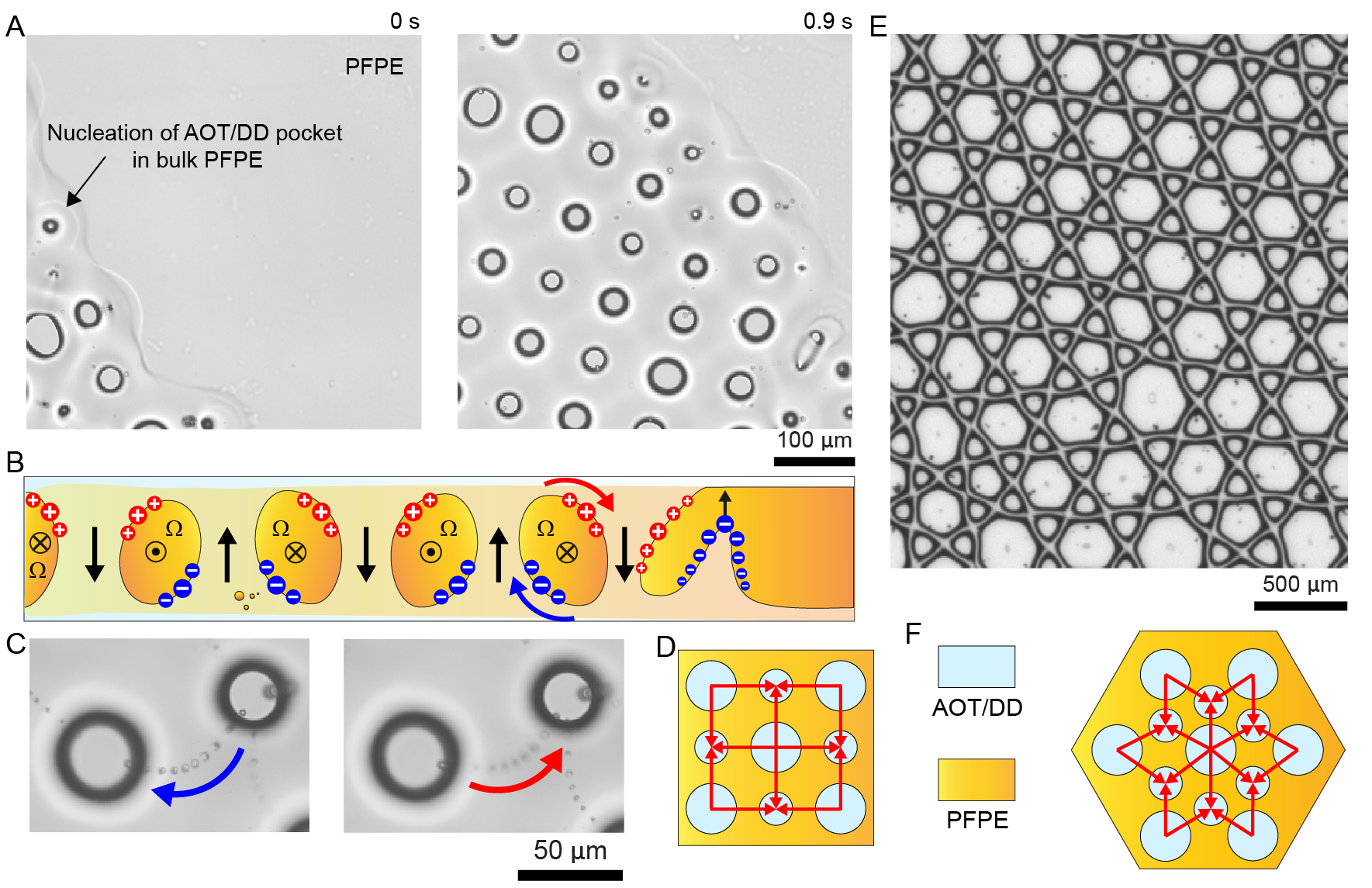}
\caption{
\textbf{Non-equilibrium fluidic lattices.}
\textbf{A,} Images of nucleation and propagation of pockets of AOT/DD away from the oil-oil interface upon application of voltage $E_0>E_c$ (Movie S4) leading to the emergence of a square lattice.
\textbf{B,} Corresponding scheme illustrating the mechanism. 
\textbf{C,} Overlaid images showing a tracer particle moving between two neighboring pockets in the square lattice (Movie S8).
\textbf{D-E,} Schemes of the two observed lattice types - square and Kagome, and  
\textbf{F,} An image of an electrohydrodynamically driven Kagome lattice (Movie S7).
}  
\label{fig:tiling2}
\end{figure}

\subsection*{Electrohydrodynamically sculpted active droplets and emulsions}
In addition to bicontinuous active filaments and dissipative lattices (Fig. 3, 4), discontinuous active droplets also emerge in the system (Fig. 5). Active droplets are most often formed from the active filaments in a process reminiscent of the Plateau-Rayleigh instability,\cite{Eggers1997} but with a fast rotating liquid filament (Fig. 5A, Movie S10). A single instability event along a filament leads to pinching of microdroplets with diameters $d$ much smaller than the cell height $h$ (Fig. 5A, B). If two instability events take place along a single filament, larger droplets are formed (Fig. 5A, B \& Supplementary Fig. S3A). The small droplets ($d < h$) are not significantly deformed by confinement or electrohydrodynamic shearing and spontaneously become self-propulsive with translational velocities up to mm/s (Movie S11). We attribute this propulsion to Quincke rotation that has been shown to lead to motility of solid colloidal particles,\cite{Bricard2013, Bricard2015, Morin2017} but now with different interdroplet interactions due to the fluidic nature of the rollers (Movie S11). 

The larger droplets ($d > h$), on the other hand, are significantly deformed by confinement and electrohydrodynamic flows, resulting in unexpected droplet shapes. For example, the large droplets can adopt polygonal shapes with up to $n$ = 12 vertices (Fig. 5C, Supplementary Fig. S4A-B). The number of vertices is always even, and the droplets appear almost stationary (non-moving and non-shape-shifting). However, as in the steady-state, interfacial patterns at low electrohydrodynamic driving (Fig. 2D), the polygonal droplets experience continuous strong internal electrohydrodynamic flows (Movie S12). In addition, the vertices of the polygons seem to correspond to the peaks of the interfacial patterns (Fig. 2F) and the bumps at the filament-filament junctions (Fig. 3D-E) where flow velocities vanish (Fig. 5D). Similarly, the edges of the polygons seem to correspond to the valleys of the interfacial patterns (Fig. 2D). Thus, a polygonal droplet with $n$ vertices possesses an apparent n-fold rotational symmetry, but when internal flows are taken into account, it has an n/2-fold rotational symmetry (Supplementary Fig. S4B), explaining why only even-sided polygons were observed. There also appears to be a critical size ($n >$ 12 for the studied system) beyond which the polygonal droplets are no longer stable (Supplementary Fig. S3B, Movie S13). 

\begin{figure}[t]
\includegraphics[width=\linewidth]{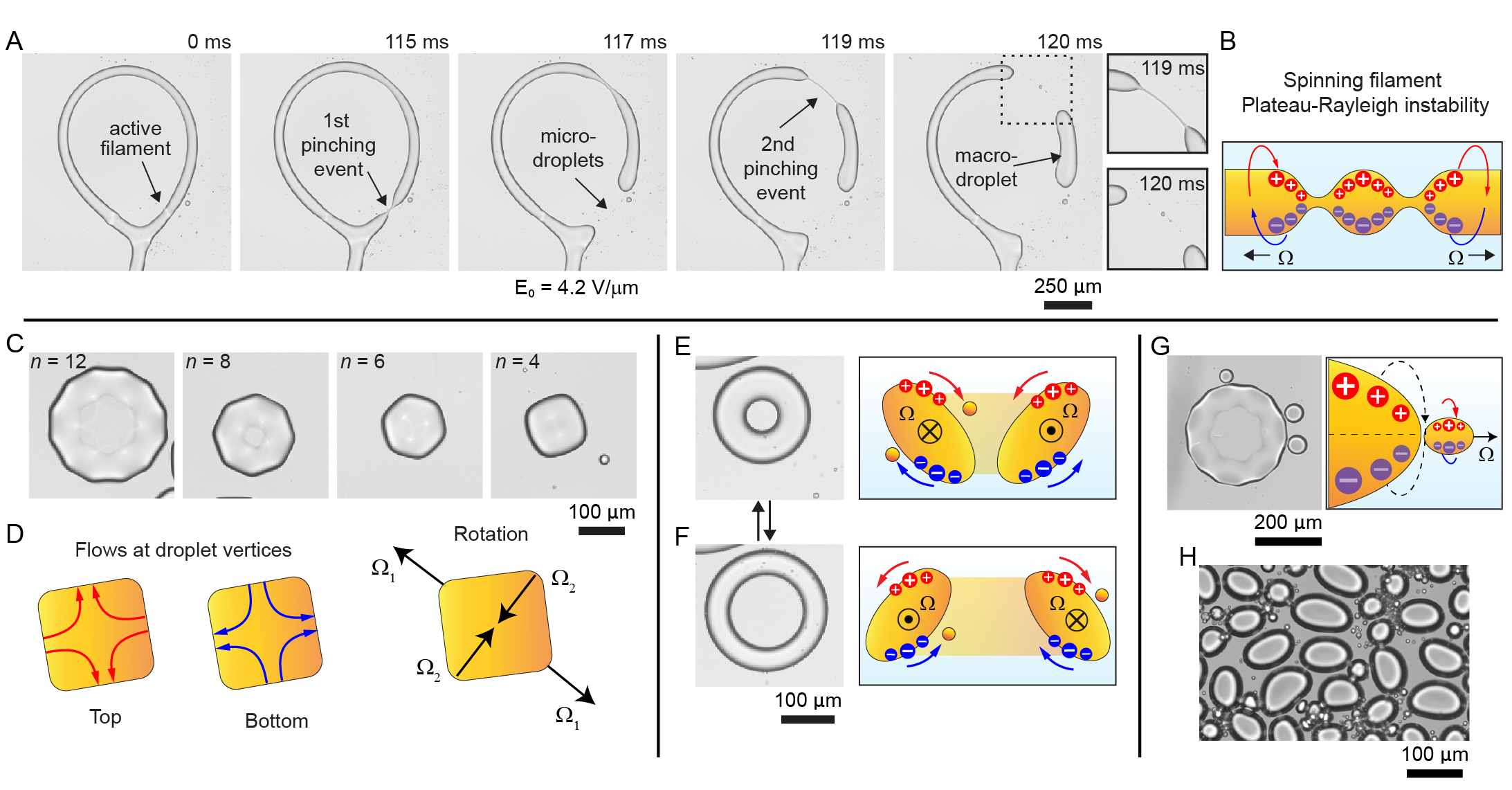}
\caption{\textbf{Emergence and dynamics of active liquid droplets, self-assemblies and emulsions.} 
\textbf{A,} Time series of microscopy images showing formation of active droplets.
\textbf{B,} Scheme of Spinning filament Plateau-Rayleigh instability in a rotating liquid filament. 
\textbf{C} Images of typical polygonal droplets. 
\textbf{D,} Scheme showing flow fields and rotational directions for the square droplet ($n=$ 4). 
\textbf{E-F,} Images of two toruses with opposing directions of rotation, and their corresponding schemes.
\textbf{G,} Images and schemes of droplet self-assemblies.
\textbf{H,} An image of a high concentration emulsion of non-coalescing active droplets (Movie S17).}
\label{fig:droplets}
\end{figure}

In addition to the polygonal droplets, another peculiar non-equilibrium droplet type is the active toroid made of a looped PFPE filament (Fig. 5E-F, Movie S14). Active toroids can rotate inwards or outwards, resulting in slightly different apparent sizes (Fig. 5E-F). This is in agreement with the observation of slightly different hole sizes in electrohydrodynamically driven lattices (Fig. 4A, E), and may originate from either gravity-induced symmetry breaking or a spontaneous symmetry breaking in the vertical direction.\cite{Vlahovska2019} The direction of rotation of a toroid can also spontaneously change in a complicated electrohydrodynamic event that starts with a Plateau-Rayleigh-like instability along the toroid (Supplementary Fig. S5, Movie S14). 

Although all of the presented active particles are liquid droplets, they do not coalesce, even in close contact. This is seen in self-assemblies of droplets, wherein smaller droplets tend to localize near vertices of the polygonal droplets (Fig. 5G, Movies S15 \& S16). The non-coalescence is strikingly evident at high droplet concentrations, and allows formation of active emulsions (Fig. 5H, Movie S17).

\subsection*{Conclusion and outlook}

In conclusion, we have shown that an unexpected diversity of organized non-equilibrium states, including interfacial patterns, active filaments, filament networks, ordered fluidic lattices, active droplets with polygonal and toroidal shapes, droplet self-assemblies and active emulsions can emerge when driving a simple biphasic planar system with non-wetting and weakly adhesive droplets out of thermodynamic equilibrium using electrohydrodynamic shearing. We foresee that this exceptionally rich dynamics of the system will open several new avenues in the fields of electrohydrodynamics,\cite{Vlahovska2019} capillary phenomena,\cite{deGennes2004} non-equilibrium pattern formation,\cite{Cross2009} dissipative self-assembly,\cite{Whitesides2002} active particle physics,\cite{Marchetti2013} and active emulsions.\cite{Weber2019, Mullol2020}




\section*{Acknowledgements}
\paragraph{Funding:}
J.V.I.T acknowledges funding from the Academy of Finland (316219) and European Research Council (803937).
\paragraph{Author Contributions:}
G.R. performed experiments and analyzed data, N.K. participated in data analysis and interpretation, J.V.I.T. conceived concepts and supervised research. All authors wrote the manuscript.
\paragraph{Competing interests:}
The authors declare no conflict of interest.
\paragraph*{Data and materials availability:}
All data is available in the manuscript or the supplementary materials.

\section*{List of Supplementary Materials}
\begin{itemize}
    \item Materials and Methods
    \item Table S1 - S3
    \item Fig S1 - S5
    \item Movie S1 - S17
\end{itemize}

Correspondence and requests for materials should be addressed to J.V.I.T. (\href{mailto:jaakko.timonen@aalto.fi}{jaakko.timonen@aalto.fi})

\section*{Supplementary Materials}
\section*{Materials and Methods}

\subsection*{Materials}
n-Dodecane (99\%, anhydrous, Acros Organics), perfluoropolyether (Krytox GPL 102-500, Chemours Performance Lubricants) docusate sodium salt (AOT, $\geq$99\%, anhydrous, Sigma-Aldrich) and glass slides coated with indium tin oxide and silver contact strips (75 x 25 x 1 mm$^3$, nominal coating thickness c.a. 350 nm, Diamonds Coatings Ltd.) were used as obtained from the manufacturer. Thermoplastic ionomer spacer film was custom made from ionomer granules (Surlyn 1702, DuPont Finland) by Muovipoli Oy (Finland). All materials were handled under ambient laboratory conditions (temperature $T\approx$ 22 $^{\circ}$C, relative humidity between 12\% and 18\%).

The leaky dielectric solution (AOT/dodecane) with a nominal concentration of 150 mM was prepared by mixing 3.34 g of AOT and 48.41 mL of dodecane. The AOT dissolved completely without any agitation in approximately 4 hours. 75 mM solution was prepared by mixing 1 mL of the 150 mM solution with 1 mL of n-Dodecane. 

\subsection*{Visualization of contact angles and adhesion}
The contact angles and adhesion of perfluoropolyether on the ITO surface was visualized using an optical goniometer (Biolin Scientific Attension Theta). A small piece of ITO-coated glass slide (c.a. 9.5 x 9.5 x 1 mm$^3$) was cut using a diamond glass-cutter and placed at the bottom of a glass cuvette (Hellma Analytics, 45 x 10 x 10 mm$^3$), with the ITO-coated surface facing upwards. The cuvette was then half-filled with the AOT/dodecane solution. Perfluoropolyether droplets were manually dispensed using a 1 mL syringe (Hanke Sass Wolf GmbH) and a needle (nominal outer diameter $\phi_{\mbox{\scriptsize{o}}} = 0.72$ mm, nominal inner diameter $\phi_{\mbox{\scriptsize{i}}} = 0.41$ mm (Hamilton Bonaduz, Kel-F hub needles)) mounted on the goniometer. 

\subsection*{Measurements of oil densities}
The densities of the 150 mM AOT in dodecane solution and perfluoropolyether (Supplementary  Table S1) were measured by aspirating 100 $\mu$L of each liquid with a positive displacement pipette (Eppendorf Multipette E3x) and measuring the mass of the aspirated liquid using an analytical balance (Ohaus Pioneer). 

\subsection*{Determination of interfacial tension}
The interfacial tension was determined using the pendant drop method, using the same setup used for visualization of contact angles and adhesion, with the exception that the needle was kept far from the ITO-coated glass slide. An image of a perfluoropolyether droplet hanging steadily was analyzed using the goniometer software (Biolin Scientific OneAttension), utilizing measured liquid densities (Supplementary Table S1).

\subsection*{Planar sample cell construction}
Each planar sample cell (Hele-Shaw cell) was constructed from two pristine ITO-coated glass slides and a Surlyn spacer. A square opening of ca. 1 cm by 1 cm was cut using a razor blade along one side of the rectangular spacer film of ca. 4 cm by 2.5 cm. The spacer film was placed between the ITO-slides and briefly melted by placing the cell on a hotplate (Fisher Scientific) at 100 $^{\circ}$C for 10 minutes, with a small weight of 20-30 grams on top of the cell. Upon cooling, the molten spacer film adheres strongly to the ITO coated glass slides, resulting in a robust cell that can hold approximately 6.5 $\mu$L of liquid in square planar confinement with height of ca. 65 $\mu$m (Supplementary Fig. 1a). Three sides of the confinement are formed by the spacer film; the fourth is open to air, and is used to fill the sample with the biphasic system.

\subsection*{Determination of the sample cell thickness}
The sample cell thickness was determined using a homemade white light interferometer constructed from a white LED (Thorlabs MWWHF2) connected to a fiber optic reflection probe bundle (Thorlabs RP22), passing the light to the sample and also collecting the reflected light and passing it to a spectrometer (Thorlabs CCS100/M). This setup was operated using software provided by the manufacturer (Thorlabs OSA). The cell height, $h$, was determined from the interference spectrum as
\begin{equation}
    h = p\frac{\lambda_p\lambda_0}{2n(\lambda_0-\lambda_p)}
    \label{cellHeight}
\end{equation}
where $\lambda_0$ is wavelength of one (arbitrary, 0-th) interference peak, $\lambda_p$ is the $p$-th peak after the 0-th peak at $\lambda_0$, and $n$ is the refractive index of the medium (air).\cite{Besseling2014} The heights of all sample cells used here (Supplementary Table S2) are averages of nine measurements per cell from different locations (Supplementary Fig. 1b). Variability of cell thickness within one cell was characterized as the standard deviation of the nine measurements, and varied from approximately 0.2 $\mu$m to 0.9 $\mu$m.

\subsection*{Measurements of oil conductivities}
Oil conductivities (Supplementary Table S1) were measured in the planar cells by pipetting a small volume of each oil into a planar cell, followed by application of a DC voltage of $U = $ 100 V and measurement of the resulting current $I$ using a sourcing electrometer (Keysight B2987A). The conductivity was calculated as $\sigma=h I / A U$, where $A$ is the area occupied by the oil in the cell, determined using image analysis (ImageJ) from an acquired image of the filled cell. The conductivity of the spacer material was assumed to be negligible.

\subsection*{Optical microscopy under electrohydrodynamic shearing}
The sample cells were filled using capillary action by placing the tip of an oil-filled pipette (Eppendorf Research plus, 10 $\mu$L) near the open side of the cell. First, 50-70\% of the cell volume was filled with the AOT/dodecane solution, then the remaining 30-50\% was filled with perfluoropolyether. 

The filled cell was observed using a modular homemade microscopy setup with transmitted light illumination, allowing the use of various magnifications and imaging speeds. Briefly, the main components of the system include optomechanical components (Thorlabs), an infinity-corrected objective lens (Nikon 1x/0.04, 5x/0.15, 10x/0.30, 20x/0.45 or 50x/0.80) and a tube lens (Thorlabs) or a finite-conjugate low-magnification zoom lens (Edmund Optics VZM 450, 0.75x - 4.5x), a USB3 camera (PointGray Grasshopper GS3-U3-51S5M-C, 2448x2048 at 75 fps) or a high-speed camera (Phantom Miro M310, 1280x800 at 3200 fps), and a collimated light source (Thorlabs MWWHLP1 LED coupled to SM2F32-A collimator). The magnification and length scale of the images was calibrated using a calibration target (Thorlabs R1L3S2P). The DC electric field was applied to the sample using the sourcing electrometer (Keysight B2987A) and hookup wires rated to 600 V (AlphaWire 2936). Wires were attached to the silver contact bars on the ITO slides with small magnets. The cameras and electrometer were operated using their manufacturers' software. See Supplementary Table S3 for a complete list of experimental conditions for each experiment presented in the article.

\subsection*{Image processing and analysis}
Native gamma and contrast settings were used for images produced by the two cameras. For images taken with the PointGray camera, the gamma is 1.0. For images taken with the Phantom camera, the gamma is 2.2. Contrast enhancement using histogram adjustment (ImageJ) was done only for images shown in Fig. 3e-f \& Supplementary Fig. 4b.

The interface elongation (Fig. 2a-b) was captured using the PointGray Grasshopper GS3-U3-51S5M-C at 50 fps. The image intensity along the axis perpendicular to the interface was determined using the Profile function of ImageJ, and by averaging over 30 neighboring lines of pixels. Full width at half maximum (FWHM) of the intensity profile was obtained with a custom MATLAB script for each time point (Supplementary Fig. 2c). Final reported values (Fig. 2b) are values averaged over time.

\pagebreak
\renewcommand\thetable{S\arabic{table}}
\setcounter{table}{0}

\begin{table}[!ht]
\centering
\renewcommand{\arraystretch}{1.3}
\begin{tabular}{|l|l|l|}
\hline
Sample no.   &  \begin{tabular}[c]{@{}l@{}}Cell height $h$ ($\mu$m) \end{tabular} & \begin{tabular}[c]{@{}l@{}} $E_{\mbox{\scriptsize{max}}}$ (V/$\mu$m) \end{tabular} \\ \hline
1.  & 64.1 $\pm$ 0.47                                      & 7.0               \\ \hline
2.  & 68.0 $\pm$ 0.61                                      & 6.6              \\ \hline
3.  & 66.1 $\pm$ 0.22                                         & 9.1             \\ \hline
4.  & 65.1 $\pm$ 0.70                                      & 5.8               \\ \hline
5.  & 36.1 $\pm$ 0.83                                         & 7.8              \\ \hline
6.  & 64.3 $\pm$ 0.88                                        & 4.7             \\ \hline
7.  & 70.9 $\pm$ 0.54                                        & 3.5              \\ \hline
\end{tabular}
\caption{List of sample cells, their average heights and standard deviations and maximum applied electric fields. See Materials and Methods section for details.}
\end{table}

\begin{table}[!ht]
\centering
\renewcommand{\arraystretch}{1.3}
\begin{tabular}{|l|l|l|l|l|l|}
\hline
Figure       & \begin{tabular}[c]{@{}l@{}} Magnification \end{tabular}& \begin{tabular}[c]{@{}l@{}}Voltage $U$\\ (V)\end{tabular} & \begin{tabular}[c]{@{}l@{}}Cell height $h$\\  ($\mu$m)\end{tabular} & \begin{tabular}[c]{@{}l@{}}E-field $E_0$\\  (V/$\mu$m)\end{tabular} & \begin{tabular}[c]{@{}l@{}}AOT/DD \\conc.  (mM)\end{tabular}\\ \hline
1b,c              & 0.75x            & 260                                                      & 66.1                                                            & 3.9       & 150                                             \\ \hline
2a (both)              & 4x            & 0, 250                                                        & 64.3                                                          & 0, 3.9                                                            & 150  \\ \hline
2c (insets), 2d              & 50x           & 300                                                      & 68.0                                                            & 4.4                                                 & 150   \\ \hline
3a (all)             & 0.75x          & 260                                                      & 66.1                                                          & 3.9                                                  & 150  \\ \hline
3d          & 10x           & 280                                                      & 64.3                                                          & 4.4                                                 & 150   \\ \hline
3e (both)             & 10x           & 370                                                      & 65.1                                                         & 5.7                                                 & 75   \\ \hline
3f (both)          & 10x           & 280                                          & 66.1                                                          & 4.2                & 150                              \\ \hline
3g          & 10x           & 270                                         & 66.1                                                          & 4.1                                 & 150            \\ \hline
3h (left)              & 1x            & 280                                                      & 65.0                                                            &     4.4                                                      & 150    \\ \hline
3h (right)             & 20x           & 200                                                      & 36.1                                                        & 5.5                                                  & 150  \\ \hline
3j  (both)            & 20x           & 200                                                      & 36.1                                                        & 5.5                                                  & 150  \\ \hline
4a (all)             & 10x           & 275                                                      & 66.1                                                          & 4.2                                                 & 150   \\ \hline
4c (left)         & 10x           & 280                                                      & 66.1                                                          & 4.2                                                 & 150   \\ \hline
4c (middle left)  & 10x           & 270                                                      & 66.1                                                          & 4.1                                                 & 150   \\ \hline
4c (middle right) & 10x           & 280                                                      & 66.1                                                          & 4.2                                                & 150   \\ \hline
4c (right)        & 10x           & 280                                                      & 66.1                                                          & 4.2                                                 & 150  \\ \hline
4e,f             & 10x           & 370                                                      & 65.1                                                        & 5.7                                                & 150    \\ \hline
4g        & 20x           & 295                                                      & 68.0                                                            & 4.3                                               & 150     \\ \hline
4h              & 50x           & 450                                                      & 64.1                                                        & 7.0                                                  & 150  \\ \hline
\end{tabular}
\caption{List of experimental parameters (magnification, voltage, cell height and electric field) for data presented in the main figures.}
\end{table}
\newpage
\begin{table}[!ht]
\centering
\renewcommand{\arraystretch}{1.3}
\begin{tabular}{|l|l|l|}
\hline
  & \begin{tabular}[c]{@{}l@{}}Perfluoropolyether (Krytox GPL 102) \end{tabular} & 150 mM AOT in dodecane \\ \hline
  \begin{tabular}[c]{@{}l@{}}Density $\rho$  (g/cm$^3$) \end{tabular}                    & 1.856                                   &0.758 \\ \hline
  \begin{tabular}[c]{@{}l@{}}Conductivity $\sigma$  (S/m) \end{tabular} & 3.3$\cdot 10^{-17}$ & 3.7$\cdot 10^{-8}$ \\ \hline
\end{tabular}
\caption{Density and conductivity of the two oils used in the study. See Methods section for details.}
\end{table}

\pagebreak
\renewcommand\thefigure{S\arabic{figure}}
\setcounter{figure}{0}

\begin{figure}[H]
\includegraphics[width=\linewidth]{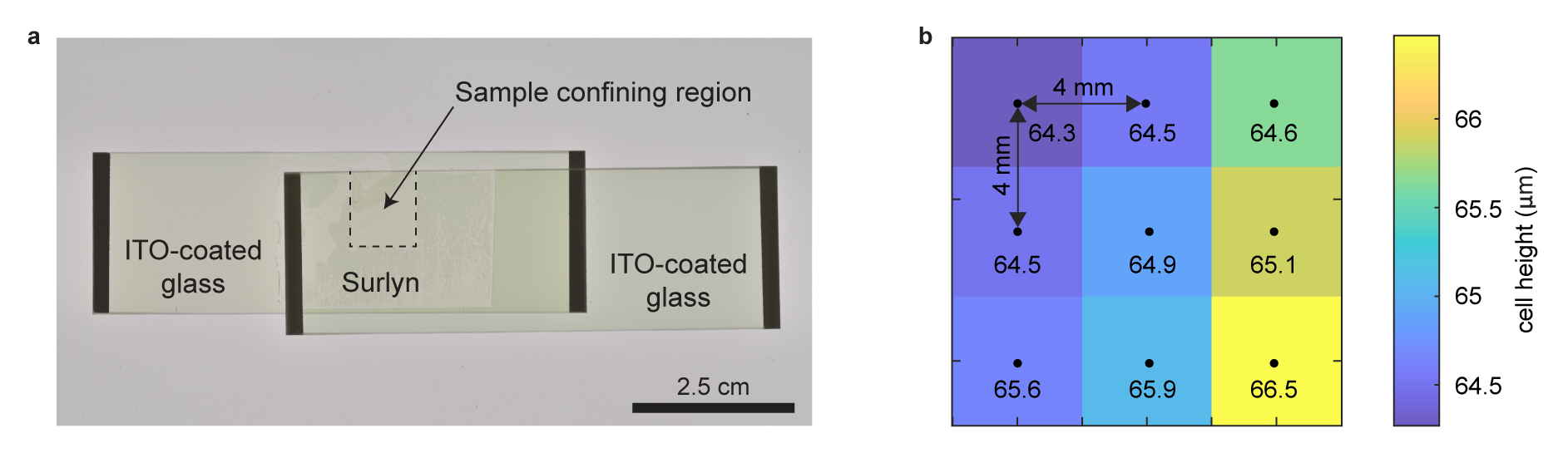}
\caption{
\textbf{Sample cell dimensions.} 
\textbf{a,} A photo of a typical sample cell used in experiments. Dashed line indicates the location of the planar confining volume. \textbf{b,} A map of cell height, \textit{h}, for a typical sample cell measured with white light interferometry. See Methods section for details of cell preparation and characterization.}
\label{fig:extOverview}
\end{figure}

\vspace{2cm}

\begin{figure}[H]
\includegraphics[width=\linewidth]{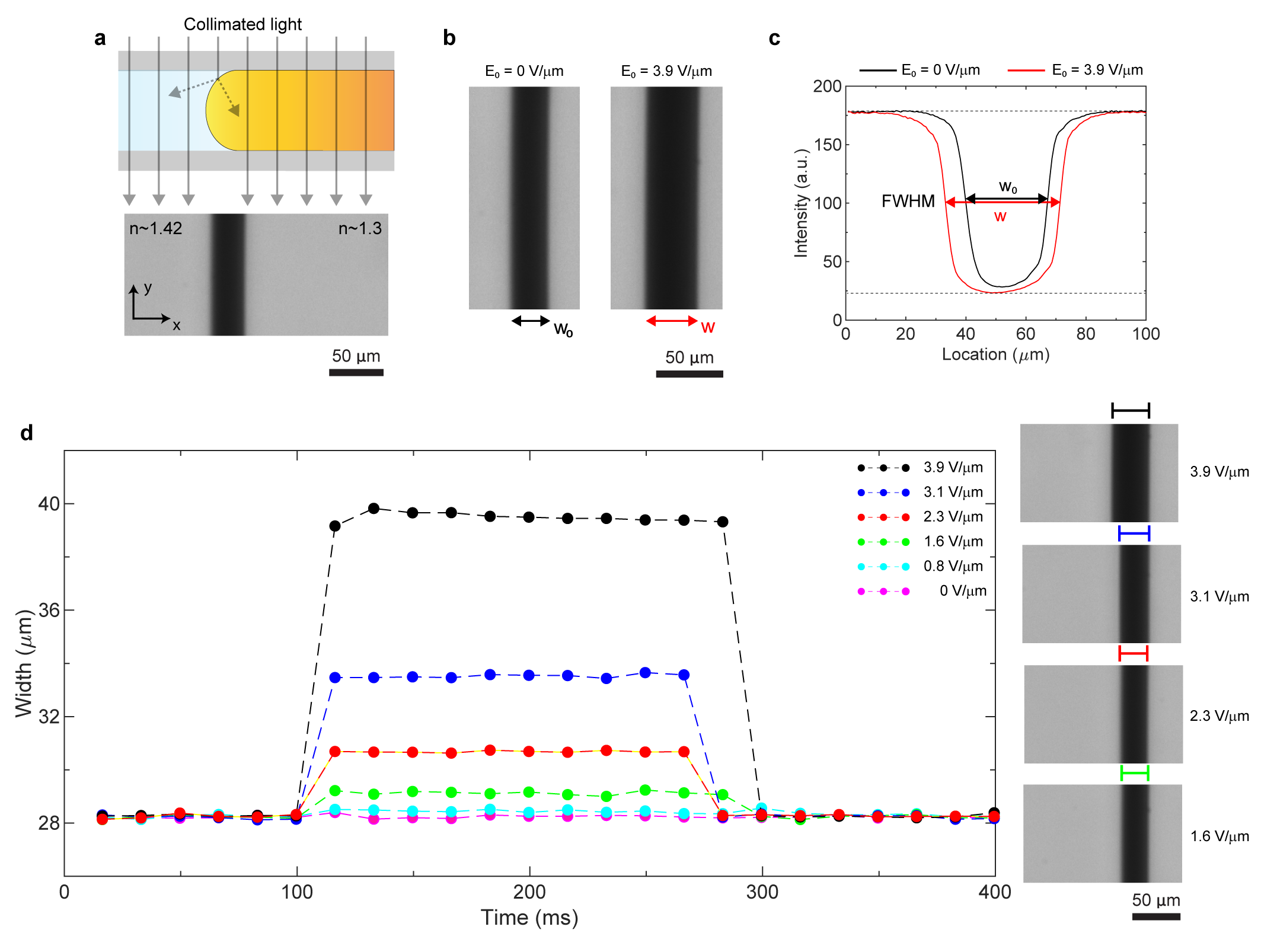}
\caption{
\textbf{Elongation of the oil-oil interface upon application of voltage across the sample cell.}
\textbf{a,} A scheme and an image showing how refraction of light at the curved oil-oil interface leads to an observable dark band in microscopy images (Video S2).
\textbf{b,} Typical image intensity profiles across the oil-oil interface, with and without an applied field.
\textbf{c,} Plot showing the width (FWHM) of the interface as a function of time for different applied voltages ranging from 0 to 250 V. For each curve, voltage was applied for 100 ms. Corresponding steady-state images at four largest voltages are shown on the right.
}
\label{fig:intfLong}
\end{figure}



\begin{figure}[H]
\includegraphics[width=\linewidth]{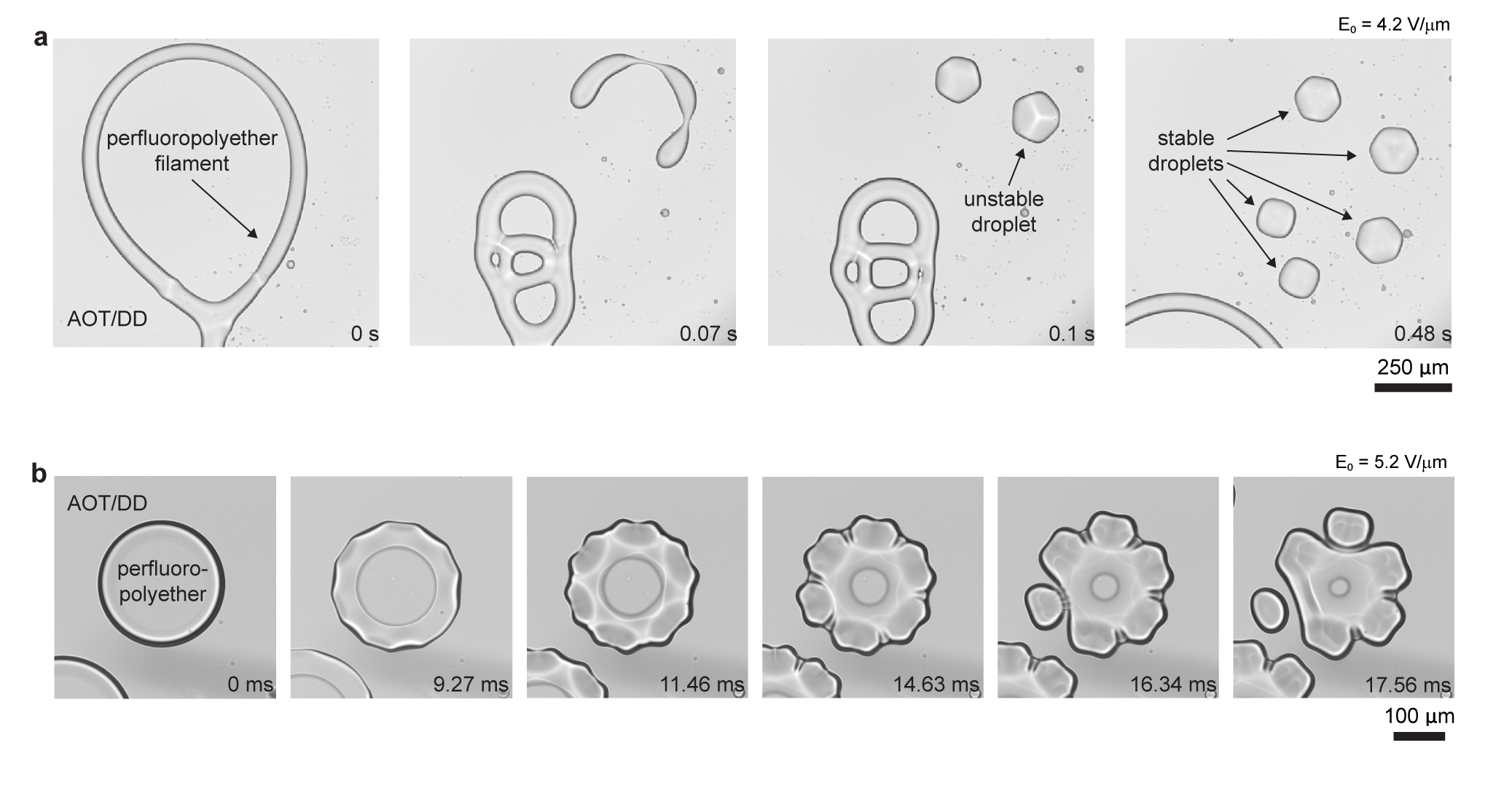}
\caption{
\textbf{Formation of active droplets}
\textbf{a,} Time series of images showing an active filament splitting several times, leading to formation of multiple small and large droplets.
\textbf{b,} Time series of images showing an unstable polygonal droplet, leading to formation of large droplets.
}
\label{fig:ActDrops1}
\end{figure}

\begin{figure}[H]
\includegraphics[width=\linewidth]{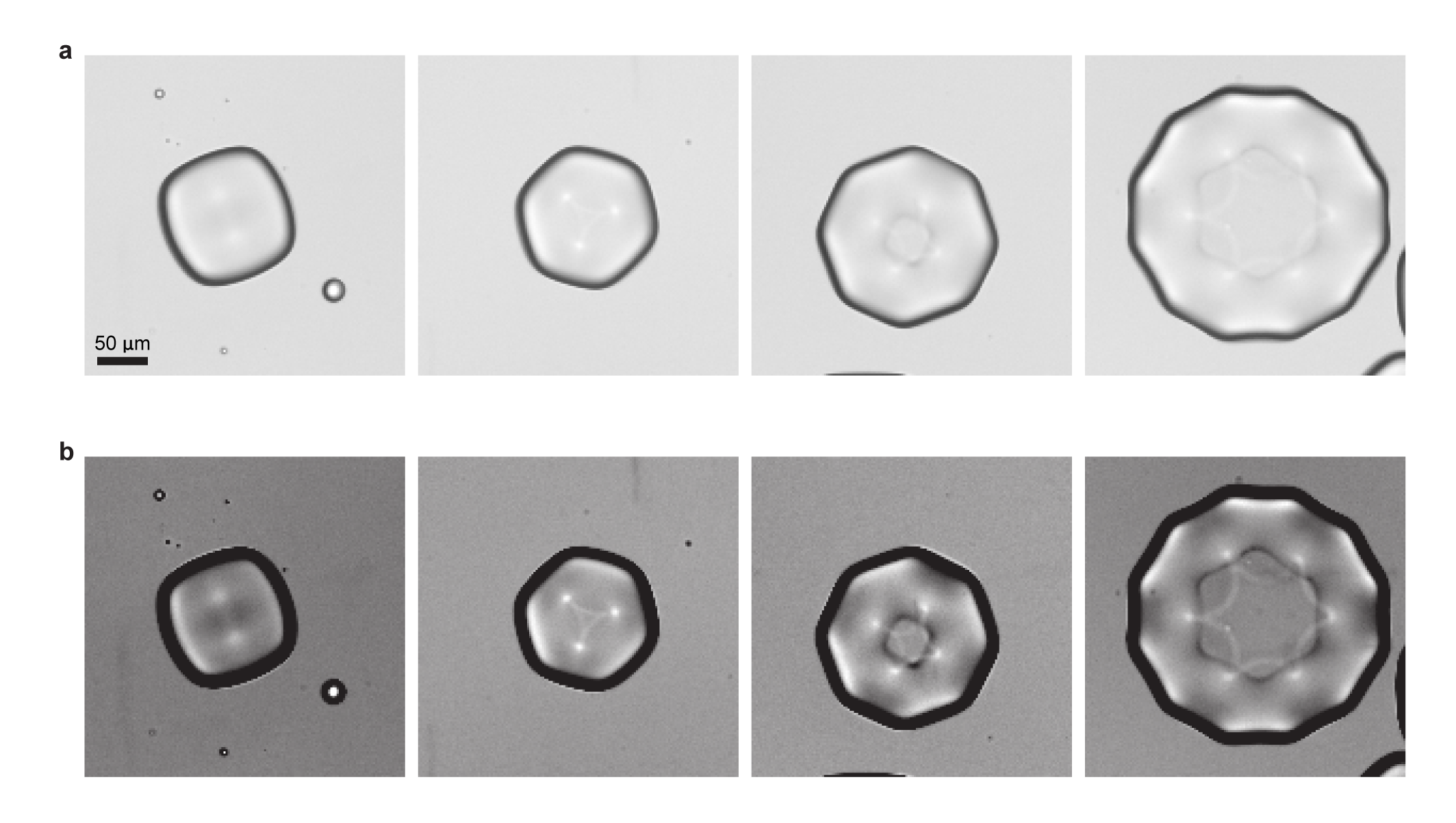}
\caption{
\textbf{Detailed visualization of polygonal droplets.}
\textbf{a,} Images of various polygonal droplets (Fig. 4e) and \textbf{b,} their contrast enhanced versions. The true 2-fold, 3-fold, 4-fold and 6-fold rotational symmetries are evident in the contrast-enhanced images.
}
\label{fig:ActDrops2}
\end{figure}

\vspace{2cm}

\begin{figure}[H]
\includegraphics[width=\linewidth]{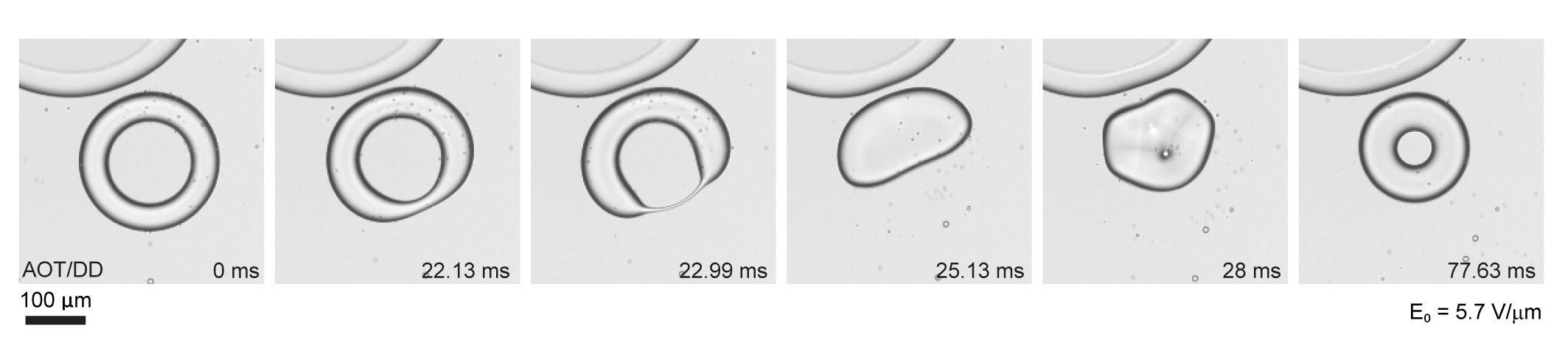}
\caption{
\textbf{Active toroid undergoing switching of the direction of rotation.}
Time series of images showing a torus reversing the rotational direction upon interaction with a neighboring droplet. Concentration of AOT/dodecane 75 mM. 
}
\label{fig:ActDrops3}
\end{figure}
\pagebreak
The movies are accessible at: \url{https://tinyurl.com/y4tmg67g}
\begin{itemize}
    \item
    \textbf{Movie S1 – Emergence of non-equilibrium states driven by electrohydrodynamic shearing} \\
    Video shows appearance of active filaments and bicontinuous fluidic lattices near the oil-oil interface. $E_0$ = 3.9 V/$\mu$m. Magnification = 0.75x. Video recorded at 60 fps, played at 30 fps.
    \item 
    \textbf{Movie S2 – Emergence of a 1D corrugation pattern (low magnification)} \\
    Video shows appearance of a typical corrugation pattern. $E_0$ = 4.4 V/$\mu$m. Magnification = 4x. Video recorded at 4100 fps, played at 30 fps.
    \item 
    \textbf{Movie S3 – Emergence of filament networks and lattices} \\
    Video shows emergence of bicontinuous filament networks and lattices. $E_0$ = 3.9 V/$\mu$m. Magnification = 0.75x. Video recorded at 60 fps, played at 30 fps.
    \item 
    \textbf{Movie S4 – Formation of holes in the perfluoropolyether drop} \\
    Video shows formation of holes in the slab-like perfluoroplyether drop. $E_0$ = 5.5 V/$\mu$m. Magnification = 20x. Video recorded at 15000 fps, played at 30 fps.
    \item 
    \textbf{Movie S5 – Rolling of active filament} \\
    Video shows an active perfluoropolyether filament and tracer particles flowing around it. $E_0$ = 4.2 V/$\mu$m. Magnification = 10x. Video recorded at 7200 fps, played at 30 fps.    
    \item 
    \textbf{Movie S6 – Lasso-like filament junction} \\
    Video shows a lasso-like filament junction. $E_0$ = 4.2 V/$\mu$m. Magnification = 10x. Video recorded at 7200 fps, played at 10 fps.    
    \item 
    \textbf{Movie S7 – Bicontinuous fluidic Kagome lattice} \\
    Video shows dynamics of a fluidic Kagome lattice with rounded hexagonal and triangular holes in a slab-like perfluoropolyether droplet. $E_0$ = 4.4 V/$\mu$m. Magnification = 1x. Video recorded at 15 fps, played at 30 fps.
    \item 
    \textbf{Movie S8 – Bicontinuous fluidic square lattice} \\
    Video shows dynamics and tracer particle movement in a square fluidic lattice. $E_0$ = 5.5 V/$\mu$m. Magnification = 50x. Video recorded at 15000 fps, played at 30 fps.    
    \item 
    \textbf{Movie S9 – Relaxation when electrohydrodynamic shearing is turned off} \\
    Video shows relaxation of complex dissipative fluidic structures to equilibrium droplets upon turning off the electrohydrodynamic shearing. $E_0$ = 4.4 V/$\mu$m. Magnification = 0.75x. Video recorded at 15 fps, played at 15 fps.
    \item 
    \textbf{Movie S10 – Formation of active droplets} \\
    Video shows formation of perfluoropolyether micro- and macrodroplets through Plateau-Rayleigh-like instability events. $E_0$ = 4.2 V/$\mu$m. Magnification = 10x. Video recorded at 2000 fps, played at 30 fps.    
   \item
   \textbf{Movie S11 – Self-propulsive active Quincke rollers)} \\
   Video shows interactions of active perfluoropolyether droplets. $E_0$ = 7.8 V/$\mu$m. Magnification = 20x. Video recorded at 4100 fps, played at 60 fps.    
   \item 
   \textbf{Movie S12 – Flows around polygonal and toroidal droplets} \\
   Video shows motion of perfluoropolyether tracer particles moving around larger perfluoropolyether drops. $E_0$ = 5.8 V/$\mu$m. Magnification = 10x. Video recorded at 15000 fps, played at 20 fps.
   \item 
   \textbf{Movie S13 – Instability of a polygonal droplet} \\ 
   Video shows formation of droplets via splitting of an unstable polygonal droplet. $E_0$ = 5.2 V/$\mu$m. Magnification = 20x. Video recorded at 4100 fps, played at 10 fps.
   \item 
   \textbf{Movie S14 – Torus switching rotation direction} \\
   Video shows a perfluoropolyether torus switching direction of rotation. $E_0$ = 5.7 V/$\mu$m. Magnification = 10x. Video recorded at 15000 fps, played at 30 fps.
   \item 
   \textbf{Movie S15 – Dissipative droplet self-assemblies 1} \\
   Video shows interaction and self-assembly of two active droplets. $E_0$ = 4.7 V/$\mu$m. Magnification = 50x. Video recorded at 4100 fps, played at 30 fps.
   \item 
   \textbf{Movie S16 – Dissipative droplet self-assemblies 2} \\
   Video shows interaction and self-assembly of multiple active droplets. $E_0$ = 4.3 V/$\mu$m. Magnification = 20x. Video recorded at 4100 fps, played at 60 fps.
   \item 
   \textbf{Movie S17 – Active emulsion} \\
   Video shows high concentration of non-coalescing active perfluoropolyether droplets, forming an active emulsion. $E_0$ = 7.0 V/$\mu$m. Magnification = 50x. Video recorded at 4100 fps, played at 30 fps.
\end{itemize}

\bibliography{References/references}

\begin{thebibliography}{10}

\bibitem{Cross2009}
M.~Cross, H.~Greenside, {\it Pattern {Formation} and {Dynamics} in
  {Nonequilibrium} {Systems}\/} (Cambridge University Press, Cambridge, 2009).

\bibitem{Marchetti2013}
M.~C. Marchetti, {\it et~al.\/}, {\it Rev. Mod. Phys.\/} {\bf 85}, 1143 (2013).

\bibitem{Fialkowski2006}
M.~Fialkowski, {\it et~al.\/}, {\it J. Phys. Chem. B\/} {\bf 110}, 2482 (2006).

\bibitem{Fodor2016}
E.~Fodor, {\it et~al.\/}, {\it Phys. Rev. Lett.\/} {\bf 117}, 038103 (2016).

\bibitem{deGennes2004}
P.-G. deGennes, F.~Brochard-Wyart, D.~Quere, {\it Capillarity and {Wetting}
  {Phenomena}: {Drops}, {Bubbles}, {Pearls}, {Waves}\/} (Springer, New York,
  2003), 2004th edn.

\bibitem{Vlahovska2019}
P.~M. Vlahovska, {\it Annu. Rev. Fluid Mech.\/} {\bf 51}, 305 (2019).

\bibitem{Hagan2016}
M.~F. Hagan, A.~Baskaran, {\it Current Opinion in Cell Biology\/} {\bf 38}, 74
  (2016).

\bibitem{Whitesides2002}
G.~M. Whitesides, B.~Grzybowski, {\it Science\/} {\bf 295}, 2418 (2002).

\bibitem{Grzybowski2000}
B.~A. Grzybowski, H.~A. Stone, G.~M. Whitesides, {\it Nature\/} {\bf 405}, 1033
  (2000).

\bibitem{Palacci2013}
J.~Palacci, S.~Sacanna, A.~P. Steinberg, D.~J. Pine, P.~M. Chaikin, {\it
  Science\/} {\bf 339}, 936 (2013).

\bibitem{Prost2015}
J.~Prost, F.~J{\"u}licher, J.-F. Joanny, {\it Nature Physics\/} {\bf 11}, 111
  (2015).

\bibitem{Nedelec1997}
F.~J. N{\'e}d{\'e}lec, T.~Surrey, A.~C. Maggs, S.~Leibler, {\it Nature\/} {\bf
  389}, 305 (1997).

\bibitem{Boekhoven2015}
J.~Boekhoven, W.~E. Hendriksen, G.~J.~M. Koper, R.~Eelkema, J.~H.~v. Esch, {\it
  Science\/} {\bf 349}, 1075 (2015).

\bibitem{vanRossum2017}
S.~A. P.~v. Rossum, M.~Tena-Solsona, J.~H.~v. Esch, R.~Eelkema, J.~Boekhoven,
  {\it Chem. Soc. Rev.\/} {\bf 46}, 5519 (2017).

\bibitem{De2018}
S.~De, R.~Klajn, {\it Advanced Materials\/} {\bf 30}, 1706750 (2018).

\bibitem{Jhawar2020}
J.~Jhawar, {\it et~al.\/}, {\it Nature Physics\/} {\bf 16}, 488 (2020).

\bibitem{Couzin2003}
I.~D. Couzin, J.~Krause, {\it Advances in the study of behaviour\/} {\bf 32}, 1
  (2003).

\bibitem{Bricard2013}
A.~Bricard, J.-B. Caussin, N.~Desreumaux, O.~Dauchot, D.~Bartolo, {\it
  Nature\/} {\bf 503}, 95 (2013).

\bibitem{Bricard2015}
A.~Bricard, {\it et~al.\/}, {\it Nature Communications\/} {\bf 6}, 7470 (2015).

\bibitem{Morin2017}
A.~Morin, N.~Desreumaux, J.-B. Caussin, D.~Bartolo, {\it Nature Physics\/} {\bf
  13}, 63 (2017).

\bibitem{Rozynek2014}
Z.~Rozynek, A.~Mikkelsen, P.~Dommersnes, J.~O. Fossum, {\it Nature
  Communications\/} {\bf 5}, 3945 (2014).

\bibitem{Dommersnes2013}
P.~Dommersnes, {\it et~al.\/}, {\it Nature Communications\/} {\bf 4}, 2066
  (2013).

\bibitem{Ouriemi2014}
M.~Ouriemi, P.~M. Vlahovska, {\it Journal of Fluid Mechanics\/} {\bf 751}, 106
  (2014).

\bibitem{Ouriemi2015}
M.~Ouriemi, P.~M. Vlahovska, {\it Langmuir\/} {\bf 31}, 6298 (2015).

\bibitem{Salipante2010}
P.~F. Salipante, P.~M. Vlahovska, {\it Physics of Fluids\/} {\bf 22}, 112110
  (2010).

\bibitem{Vlahovska2016}
P.~M. Vlahovska, {\it Phys. Rev. Fluids\/} {\bf 1}, 060504 (2016).

\bibitem{Brosseau2017}
Q.~Brosseau, P.~M. Vlahovska, {\it Phys. Rev. Lett.\/} {\bf 119}, 034501
  (2017).

\bibitem{Varshney2012}
A.~Varshney, S.~Ghosh, S.~Bhattacharya, A.~Yethiraj, {\it Scientific Reports\/}
  {\bf 2}, 738 (2012).

\bibitem{Tadavani2016}
S.~K. Tadavani, J.~R. Munroe, A.~Yethiraj, {\it Soft Matter\/} {\bf 12}, 9246
  (2016).

\bibitem{Timonen2013}
J.~V.~I. Timonen, M.~Latikka, L.~Leibler, R.~H.~A. Ras, O.~Ikkala, {\it
  Science\/} {\bf 341}, 253 (2013).

\bibitem{Taylor1966}
G.~I. Taylor, A.~D. McEwan, L.~N.~J. de~Jong, {\it Proceedings of the Royal
  Society of London. Series A. Mathematical and Physical Sciences\/} {\bf 291},
  159 (1966).

\bibitem{Westra2003}
M.-T. Westra, D.~J. Binks, W.~V.~D. Water, {\it Journal of Fluid Mechanics\/}
  {\bf 496}, 1 (2003).

\bibitem{Rosensweig2014}
R.~E. Rosensweig, {\it Ferrohydrodynamics\/} (Dover Publications, New York,
  2014).

\bibitem{Quincke1896}
G.~Quincke, {\it Annalen der Physik\/} {\bf 295}, 417 (1896).

\bibitem{Pradillo2019}
G.~E. Pradillo, H.~Karani, P.~M. Vlahovska, {\it Soft Matter\/} {\bf 15}, 6564
  (2019).

\bibitem{Eggers1997}
J.~Eggers, {\it Rev. Mod. Phys.\/} {\bf 69}, 865 (1997).

\bibitem{Weber2019}
C.~A. Weber, D.~Zwicker, F.~J{\"u}licher, C.~F. Joanny, {\it Reports on
  Progress in Physics\/} {\bf 82}, 064601 (2019).

\bibitem{Mullol2020}
J.~Ign{\'e}s-Mullol, F.~Sagu{\'e}s, {\it Current Opinion in Colloid and
  Interface Science\/} {\bf 49}, 16 (2020).

\bibitem{Besseling2014}
T.~H. Besseling, J.~Jose, A.~V. Blaaderen, {\it Journal of Microscopy\/} {\bf
  257}, 142 (2015).

\end{thebibliography}

\end{document}